\documentclass[12pt,a4paper]{article}
\usepackage{a4wide}
\usepackage{latexsym}
\usepackage{epsf}
\usepackage{amssymb}
\begin{document}
\def\be{\begin{equation}}
\def\ee{\end{equation}}
\def\bea{\begin{eqnarray}}
\def\eea{\end{eqnarray}}

\def\pd{\partial}
\def\a{\alpha}
\def\b{\beta}
\def\g{\gamma}
\def\d{\delta}
\def\m{\mu}
\def\n{\nu}
\def\t{\tau} 
\def\l{\lambda}
\def\s{\sigma}
\def\e{\epsilon}
\def\scri{\mathcal{J}}
\def\cM{\mathcal{M}}
\def\tcM{\tilde{\mathcal{M}}}
\def\RR{\mathbb{R}}
\def\CC{\mathbb{C}}
%%%%%%%%%%%%%%%%%%%%%%%%%%%%%%%%%%%%%%%%%%%%%%%%%%%%%%%%%%%%%%%%%%%%%%

\hyphenation{re-pa-ra-me-tri-za-tion}
\hyphenation{trans-for-ma-tions}

%%%%%%%%%%%%%%%%%%%%%%%%%%%%%%%%%%%%%%%%%%%%%%%%%%%%%%%%%%%%%%%%%%%%%%

\begin{flushright}
IFT-UAM/CSIC-00-33\\
hep-th/0011105\\
\end{flushright}

\vspace{1cm}

\begin{center}

{\bf\Large Non-Critical Poincar\'e Invariant Bosonic String Backgrounds and
Closed String Tachyons}

\vspace{.5cm}

{\bf Enrique \'Alvarez}
\footnote{E-mail: {\tt enrique.alvarez@uam.es}}
{\bf C\'esar G\'omez}
\footnote{E-mail: {\tt cesar.gomez@uam.es  }}
{\bf and Lorenzo Hern\'andez }
\footnote{E-mail: {\tt Lorenzo.Hernandez@uam.es }} \\
\vspace{.3cm}

\vskip 0.4cm

\ {\it Departamento de F\'{\i}sica Te\'orica, C-XI,
  Universidad Aut\'onoma de Madrid \\
  E-28049-Madrid, Spain \\and\\ 
Instituto de F\'{\i}sica Te\'orica, C-XVI,
  Universidad Aut\'onoma de Madrid \\
  E-28049-Madrid, Spain}\footnote{Unidad de Investigaci\'on Asociada
  al Centro de F\'{\i}sica Miguel Catal\'an (C.S.I.C.)}

\vskip 1cm

%%%%%%%%%%%%%%%%%%%%%%%%%%%%%%%%%%%%%%%%%%%%%%%%%%%%%%%%%%%%%%%%%%%%%%

{\bf Abstract}

\end{center}
A new family of non critical bosonic string backgrounds in arbitrary
space time dimension $D$ and with $ISO(1,D-2)$ Poincar\'e invariance are
presented. The metric warping factor and dilaton agree asymptotically
with the linear dilaton background. The closed 
string tachyon equation of motion enjoys, in
the linear approximation, an exact solution of ``kink'' type
interpolating between different expectation values. A renormalization group
flow interpretation ,based on a closed string tachyon potential of type
$-T^{2}e^{-T}$, is suggested.

%%%%%%%%%%%%%%%%%%%%%%%%%%%%%%%%%%%%%%%%%%%%%%%%%%%%%%%%%%%%%%%%%%%%%%

\begin{quote}

\end{quote}

%%%%%%%%%%%%%%%%%%%%%%%%%%%%%%%%%%%%%%%%%%%%%%%%%%%%%%%%%%%%%%%%%%%%%%

\newpage
%%%%%%%%%%%%%%%%%%%%%%%%%%%%%%%%%%%%%%%%%%%%%%%%%%%%%%%%%%%%%%%%%%%%%%

\setcounter{page}{1}
\setcounter{footnote}{1}
%\renewcommand{\theequation}{\thesection.\arabic{equation}}
%%%%%%%%%%%%%%%%%%%%%%%%%%%%%%%%%%%%%%%%%%%%%%%%%%%%%%%%%%%%
\section{Introduction and Summary}
%%%%%%%%%%%%%%%%%%%%%%%%%%%%%%%%%%%%%%%%%%%%%%%%%%%%%%%%%%%%%%%

In order for strings to behave as consistent quantum systems, a mimimal 
requirement seems to be the absence of anomalies.
\par
World sheet conformal invariance at tree level in the string perturbative
expansion leads to vanishing sigma model beta functions. Denoting $D$ the
dimension of the target space-time the relevant beta functions 
for the bosonic string are

\bea\label{beta}
\beta^{g}_{AB}&&=\a'R_{AB}+ 2 \a'\nabla_A\nabla_B \Phi -\frac{\a'}{4}H_{ACD}
H_{B}^{CD}+ o(\a'^2)\nonumber\\
\beta^{b}_{AB}&&= -\frac{\a'}{2}\nabla^D H_{DAB}+ \a'\nabla^D\Phi H_{DAB}
+ o(\a'^2)\nonumber\\
\beta^{\Phi}&&=\frac{D-26}{6}-\frac{\a'}{2}\nabla^2 \Phi + 
\a'\nabla_D\Phi\nabla^D\Phi -\frac{\a'}{24}H_{ABC}H^{ABC}+ o(\a'^2)
\eea
 
The simplest solution is of course the critical string $D=26$ in flat
Minkowski space time with constant dilaton and vanishing antisymmetric
tensor. Another well known solution in $D\neq 26$ 
is the linear dilaton background \cite{myers}
\be
\Phi = q_{\mu}X^{\mu}
\ee
in flat $D$ dimensional Minkowski space-time, with

\be
q_{\mu}q^{\mu} = \frac{26-D}{6\alpha'}
\ee
The linear dilaton background solution is exact to all orders in $\alpha'$ 
and defines a good conformal field theory. Moreover for $D$ smaller or equal
to
two this background is tachyon free in the linear approximation, and is 
believed to develop a tachyonic 
barrier that prevents strings to enter in the string strong coupling region.
\par
Motivated by the {\em curved Liouville} approach of Polyakov \cite{polyakov} 
to non critical
backgrounds we will present in this letter a new family of solutions
to the sigma model beta function equations, at first order in $\alpha '$, for
generic values of $D$ and with $ISO(1,D-2)$ Poincare invariance
\be\label{metric}
ds^2= a(r)d\vec{x}_{1,D-2}^2 + dr^2 
\ee
Backgrounds of this type are typical in the holographic context once
we identify the coordinate $r$ with the holographic direction or
in curved Liouville if we interpret $r$ as the extra Liouville direction.

\begin{figure}
\begin{center}
\leavevmode
\epsfxsize=6cm
\epsffile{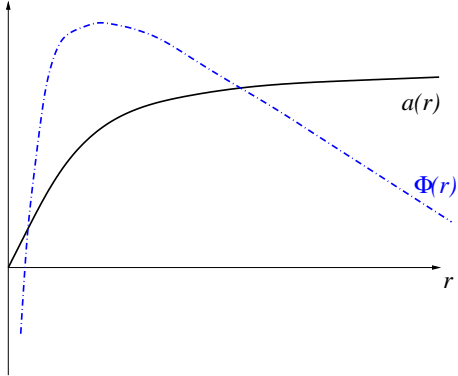}
\caption{\it Warp factor and dilaton v.s. holographic coordinate. In the $r\rightarrow \infty$ linear dilaton behavior in flat Minkowski
space time can be noticed.}
\label{fig1}
\end{center}
\end{figure}

\par
To be specific, we shall find ( see Figure 1) that in $D=5$ spacetime dimensions 
(corresponding to one holographic plus the ordinary four minkowskian ones),
the warp factor is:
\be\label{warp}
a(r) = th(\sqrt{\frac{21}{6\a^{\prime}}}r)
\ee
and the dilaton
\be\label{dil}
\Phi = \frac{1}{2} \log \left[th(\sqrt{\frac{21}{6\a^{\prime}} }r) \frac{1}
{cosh^2 (\sqrt{\frac{21}{6\a^{\prime}} }r)}\right]
\ee
Notice from equations (\ref{warp}) and (\ref{dil}) that this solution
in the asymptotic region $r \rightarrow \infty$ coincides 
with flat Minkowski space-time 
with the appropiated linear dilaton background. At $r=0$ the warp factor
goes to zero inducing a naked singularity. Close to the origin we reproduce
the metric of reference \cite{ag} with a logaritmic behavior for the dilaton.
Moreover in the limit $\alpha' \rightarrow 0$ the metric becomes flat Minkowski except
in a neighborhood of the origin \footnote{This solution then seems to embody in a natural
way stringy corrections to linear dilaton background}.

\par

%%%%%%%%FIGURA 2%%%%%%%%%%%%%%
\begin{figure}
\begin{center}
\leavevmode
\epsfxsize=6cm
\epsffile{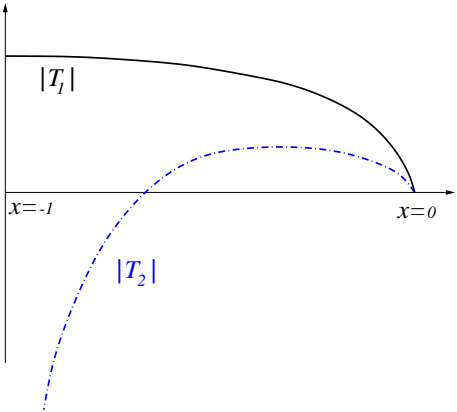}
\caption{\it The solid line represents the kink solution interpolating
between $T=0$ at $x=0, (r= \infty)$ and $T = cte$ at $x=-1 ,( r=0)$. The
variable $x$ is defined in equation (\ref{var}).}
\label{fig2}
\end{center}
\end{figure}

\par

With respect to the problem of the  tachyon
stability we study the tachyon equation of motion in the linear approximation.
The equation of motion becomes Riemann's equation with three singular points.
In the asymptotic region  $r \rightarrow \infty$ we find the familiar $c=1$ barrier
for linear dilaton background; namely for $D>2$
the tachyon goes to zero oscillating. At the origin the exact solution, preserving
energy conservation,
goes to a constant. Thus this solution
can be temptatively interpreted as a kind of
tachyonic kink ( see Fig 2) interpolating
- in the sense of a renormalization group flow- 
the unstable linear dilaton asymptotic regime with
vanishing vacuum expectation value of the tachyon and 
a stable background characterized by a non vanishing tachyon 
vacuum expectation value. This interpretation could be probably supported by
a closed string tachyon potential of type $V(T)= - e^{-T} T^{2}$ as recently
suggested in reference \cite{tseytlin}. 
In the rest of this paper we will present the technical details of
our analysis.

%%%%%%%%%%%%%%%%%%%%%%%%%%%%%%%%%%%%%%%%%%%%%%%%%%%%%%%
\section{A new family of non-critical bosonic string backgrounds}
%%%%%%%%%%%%%%%%%%%%%%%%%%%%%%%%%%%%%%%%%%%%%%%%%%%%%%%%%%
For vanishing Kalb-Ramond field the beta function equation $\beta^{g} =0$
for the Poincare invariant metric (\ref{metric}) reads:
\bea\label{beta1}
\frac{2 a a'' +(d-2)(a')^2}{4a}\eta_{\m\n}&&=a'\Phi'\eta_{\m\n}\nonumber\\
\frac{d}{4}\frac{2a a''-(a')^2}{a^2} &&= 2 \Phi''
\eea
for $d=D-1$.
The dilaton solving the first of these equations is given by
\be
\Phi = \frac{1}{2}\ log \, a'+\frac{d-2}{4}log\, a
\ee
and the general solution for the metric factor would be obtained
by integrating the following equation:
\be
\frac{da}{dr}=c_1 a^{1+\frac{\sqrt{d}}{2}}+c_2 a^{1-\frac{\sqrt{d}}{2}}
\ee
There are now several possibilities:
\begin{itemize}
\item $c_1c_2<0$.
The solution then reduces to:
\be
a(r) = \left[\frac{\sqrt{-c_1c_2}}{c_1}th\left(\sqrt{-c_1c_2 }\frac{\sqrt{d}}{2}(r+c_3)\right) \right]^{\frac{2}{\sqrt{d}}}
\ee
\item $c_1c_2>0$.
The solution reads :
\be
a(r) = \left[\frac{\sqrt{c_1c_2}}{c_1}tg \left( \sqrt{c_1c_2 }\frac{\sqrt{d}}{2}(r+c_3)\right) \right]^{\frac{2}{\sqrt{d}}}
\ee
\item $c_2=0$.
The solution is then:
\be
a(r)=(r+c_3)^{-\frac{2}{\sqrt{d}}}
\ee
(This is the dual $r \to \frac{1}{r}$ of
the solution presented in \cite{ag}).
\item $c_1=0$.
This leads to:
\be
a(r)=(r+c_3)^{\frac{2}{\sqrt{d}}}
\ee

(This is the confining background introduced in \cite{ag}).
\item $c_1=c_2=0$.
This is the linear dilaton solution:
\bea
a&&=a_0\nonumber\\
\Phi&&=c_3+c_4 r
\eea
with $c_3,c_4$ arbitrary constants. Notice that in this case equations
(\ref{beta1}) imply $\Phi'' =0$.

\end{itemize}

\par 

The most important thing about this family of solutions is that they
are non-critical; that is, if we compute the beta function of the dilaton
(which is guaranteed to be a constant by virtue of the first two equations
and Bianchi identities for the spacetime curvature,) it so happens that
\be
c_1c_2 = \frac{4}{6 \alpha '}(1 - \frac{25}{d})
\ee
This clearly means that if $c_1c_2<0$, there is a solution in the family
for any $D<26$; whereas for $c_1c_2>0$ the solution exists for any $D>26$.
\par
It is also worth remarking that as soon as one of the two arbitrary constants
vanish, the solution must necessarily become critical (that is, with $D=26$).
\par
Let us now briefly describe some properties of the solutions in the 
physically most interesting case, namely for $D < 26$. The behavior
in the asymptotic region $r\rightarrow \infty$ for the metric factor $a$
is
\be
a \rightarrow \bar{a}=(\frac{-c_{2}}{c_{1}})^{\frac{1}{\sqrt{d}}}
\ee
whereas the dilaton yields
\be
\Phi \rightarrow -\sqrt{\frac{25-d}{6 \alpha'}} r
\ee
where we easily recognize the precise behavior of the linear dilaton
solution in flat space-time. 
\par
In the region $r \rightarrow 0$ we get instead
\be
a \rightarrow \left( {\frac{c_2 \sqrt{d} }{2}}r \right)^{\frac{2}{\sqrt{d}}}
\ee
and the dilaton goes as
\be
\Phi \rightarrow (\sqrt{d} -1) \log r 
\ee
\par
Notice that all these solutions are singular at $r=0$. For instance the scalar curvature in d=4 is given by
\be
R =  {-c_1 c_2}\frac{(a^2-1)(9a^2 -1)}{a^2}  
\ee
Given the fact that this is the {\em general}\, solution with our ansatz,
this means that the existence of the naked singularity is somewhat embodied
in the ansatz itself.
\par
A simple generalization of this background
(\ref{metric}) can be obtained by adding
a certain number $d'$ of extra flat dimensions $(D=d+d'+1)$: 
\be
ds^2= a(r)d\vec{x}_{1,d}^2 + dr^2 + d\vec{y}_{d'}
\ee
The only change in this case is on the values of the constants
$c_1,c_2$ that are now fixed by the relation
\be
\frac{dc_1c_2}{4} = \frac{(d+d'-25)}{6 \alpha'}
\ee

\section{Closed String Tachyons}
%%%%%%%%%%%%%%%%%%%%%%%%%%%%%%%%%%%%%%%%%%%%%%%%%%%%%%%%%%%
%%%%%%%%%%%%%%%%%%%%%%%%%%%%%%%%%%%%%%%%%%%%%%%%%%%%%%%%%%%%%%%%%%%%
Bosonic string backgrounds suffer from an intrinsic source for instability
(which sometimes has been taken as an indication of the existence
of another, physically interesting and energetically favoured  vacuum
\cite{englert}), namely, a tachyonic scalar excitation in the
closed string sector with
\be
m^2 = -\frac{4}{\a'}. 
\ee
It was first suggested by Polyakov \cite{polyakov} that in a metric
of type (\ref{metric}) the tachyon unstability could be tamed 
leading to a peaceful condensation. In addition we know that in a non trivial
space time - AdS is a good example \cite{bf} - normalizability and
energy conservation imposses strong restrictions on the allowed 
spectrum of masses. Following this last approach the first thing we will
do would be to derive from energy conservation and normalizability, the
boundary conditions on field configurations. Next we will consider -
in the linear approximation - the
exact solution to the tachyon equation of motion satisfying these boundary
conditions. Finally we will interpret the solution as a kink describing tachyon
condensation.
\par
%%%%%%%%%%%%%%%%%%%%%%%%%%%%%%%%%%%%%%%%%%%%%%%%%%%%%%%%%%%%%%
\subsection{Boundary Conditions}
\subsubsection{Normalization}
%%%%%%%%%%%%%%%%%%%%%%%%%%%%%%%%%%%%%%%%%%%%%%%%%%%%%%%%%%%
First of all, if the scalar is complex, there is a conserved number current:
\be
N^A\equiv e^{-2\Phi}((f_2)^{*} \pd^A f_1 -f_1 \pd^A (f_2)^{*})
\ee
which gives rise to a conserved particle number
\be
N\equiv \int d^{d-1}x dr \sqrt{g} e^{-2\Phi}(f^{*} \pd^0 f -f \pd^0 f^{*})
\ee
when this integral converges. The $a$-dependent factors in the measure
behave as:
\be
N\sim a^{(d-1)/2}\frac{1}{a' a^{\frac{d-2}{2}}}
\ee
\par
In the generic case, for $c_1c_2<0$ this behaves for  $a\sim 0$
as:
\be
N\sim a^{\frac{\sqrt{d}-1}{2}}
\ee
On the other hand, when $a$ reached its asymptotic value, 
$a\rightarrow \bar{a}$,
\be
N\sim \frac{1}{\bar{a}-a}
\ee
This last behaviour selects functions that vanish as a power when 
$a\rightarrow \bar{a}$.
That is,
\be
lim_{a\rightarrow \bar{a}}f = (\bar{a}-a)^{\delta}
\ee
where $\delta >\frac{1}{2}$.

%%%%%%%%%%%%%%%%%%%%%%%%%%%%%%%%%%%%%%%%%%%%%%%%%%%%%%%%%%%%
\subsubsection{The definition of Conserved Energy}
%%%%%%%%%%%%%%%%%%%%%%%%%%%%%%%%%%%%%%%%%%%%%%%%%%%%%%%%%%%%%%

Given an arbitrary scalar field with
energy-momentum tensor $T_{AB}$, there is a {\em Killing energy} current,
namely
\be
j^{A}\equiv T^{AB} k_{B}
\ee
which is covariantly conserved, 

i.e.
\be\label{div}
\nabla_{A}j^{A} = 0
\ee
In our case, the Killing corresponding to temporal translations is:
\be
k\equiv \frac{\pd}{\pd t}
\ee
Applying Stokes'theorem to the integral over the $(d+1)$-dimensional 
region $\mathcal{R}$ defined in terms of a large distance $L$ by
 $ t_0<t< t_1$,and by $-L<x^i<L$,($i,j,\ldots =1,\ldots d-1$), $0<r<L$,
 of the $(d+1)$-form proportional
to the first member of (\ref{div}), i.e. of $d* j$, where $j$ is 
the one-form dual to
the current, $j\equiv j_{A}dx^{A}$, we get
\be
0 = \int_{\mathcal{M}} d*j = \int_{\pd \mathcal{M}} *j = E(t_2)-E(t_1) +
 \Delta E
\ee
where
\be
E(t)\equiv \int d^{(d-1)} x dr \sqrt{g}j^0(t,x^i,r)
\ee
and
the flux over the boundary $\pd M$,
\be
\Delta E = (\int_{x^i =L}-\int_{x^i = -L})(\sqrt{g}) 
(\epsilon_{i i_1 \ldots i_{d-2}}
dt\wedge dx^{i_1}\wedge\ldots\wedge dx^{i_{d-2}}\wedge dr) j^i +
(\int_{r=L}- \int_{r=0}) d^{d}x (\sqrt{g}) j^{d}.
\ee
Only when the physical boundary conditions are such that 
\be
\Delta E =0
\ee
there is actual energy conservation,
\be
\frac{dE}{dt}=0
\ee
For a scalar field of mass $m$ it is readily found that
\be
T_{AB} = \frac{1}{2}e^{-2\Phi}[\nabla_A \phi\nabla_B \phi - \frac{1}{2}g_{AB}
(g^{CD}\nabla_C \phi\nabla_D \phi - m^2 \phi^2)]
\ee
so that the $(d+1)$-dimensional energy is:
\be
E[\phi]\equiv\frac{1}{4}\int d^{(d-1)}x dr \sqrt{g}T^{0}_B k^B,
\ee
i.e.,denoting the scalar perturbation by $T$,
\be\label{E}
E = \int dr d^{d-1}x \frac{1}{\sqrt{a} a'} \left[ \dot{T}^2 + (\nabla_{d-1}T)^2
+a ((\pd_r T)^2+ m^2 T^2)\right]
\ee
\par
In all cases the important thing is the behaviour of the flux at the 
singular boundary, which in turn is dominated by the behaviour of
$j^d$, when $a\rightarrow 0$, namely
\be\label{tacc}
j^d\sim - \frac{a}{2} \frac{1}{a'a^{\frac{d-2}{2}}} \frac{1}{a}\pd_r T\dot{T}=
- \frac{1}{2} a^{1-\frac{d}{2}} T'\dot{T}
\ee

%%%%%%%%%%%%%%%%%%%%%%%%%%%%%%%%%%%%%%%%%%%%%%%%%%%%%%%%%%%%
\subsubsection{Positive-definiteness}
%%%%%%%%%%%%%%%%%%%%%%%%%%%%%%%%%%%%%%%%%%%%%%%%%%%%%%%%%%
Combining the results of the preceding two paragraphs we can read the 
conditions for the scalar perturbations to be normalizable, as well as enjoy
a conserved energy, that is,
\bea\label{funcional}
f&&\sim_{a=0} 0(a^{-u})\nonumber\\
f&&\sim_{a=\bar{a}}(\bar{a}-a)^v
\eea
with $0<u<1$ and $v>\frac{1}{2}$.
\par 
If we want moreover the now well defined energy to be finite it is 
clearly necessary (owing to the $a^{\prime}$ factor in the denominator)
 that when
$r\rightarrow \infty$ 
\be\label{mu}
T\sim e^{- \mu^2 r}
\ee
with 
\be\label{finita}
\mu^2 > \sqrt{\frac{25-d}{6 \alpha'}}
\ee
which is much faster that required by (\ref{funcional}).We can define 
another functional space in which energy is not only conserved, but
 also finite, and this space is essentially characterized by
the behavior (\ref{finita}).
\par

It is then plain that in the conserved-energy functional space 
defined by (\ref{funcional})
there are fluctuations with arbitarily large  negative energy,
namely all those that at infinity behave as (\ref{mu})
with $|\mu|< |m|$. 
In the more restrictive finite-energy space (\ref{finita}), this behavior 
still persists as long as $d>1$. This is the same bound found for the
linear dilaton background.

%%%%%%%%%%%%%%%%%%%%%%%%%%%%%%%%%%%%%%%%%%%%%%%%%%%%%%%%%%%
\subsection{The tachyon equation}
%%%%%%%%%%%%%%%%%%%%%%%%%%%%%%%%%%%%%%%%%%%%%%%%%%%%%%%
Let us now examine the space of solutions of
the linearized equation of motion and check whether on shell there
are elements of our functional space of well-defined fluctuations.
\par
We shall write the wave equations (following the conventions of
\cite{polchinski}) as:
\be\label{wave}
\nabla^2 T - 2 \nabla^A \Phi \nabla_A T - m^2 T = 0
\ee
For the Poincar\'e invariant ansatz the dilaton is given by
\be
\Phi=\frac{1}{2}log a' + \frac{d-2}{4}log a
\ee
where $a'\equiv\frac{da}{dr}$. 
\par
Variables can be separated in the wave equation (\ref{wave}) by means of:
\be
T(x^A) \equiv T(a)\Theta (x^{\m})
\ee
with the result:
\bea\label{eqa}
\Box_d \Theta && = \lambda \Theta\nonumber\\
T'' + \frac{1}{a}T'&&= (\frac{a m^2 -\lambda }{a (a')^2})T
\eea

(where we have used $\frac{dT}{dr} = \frac{dT}{da}a'\equiv T' a'$), and 
$\lambda$ is the separation constant.
\par
Now, for a plane wave $\Theta = e^{i kx}\equiv 
e^{-i(\omega t - \vec{k}\vec{x})}$, the first (Minkowskian) equation reads
\be
\omega^2 -\vec{k}^2 =\lambda
\ee
Thus, in order to be able to build wave packets of arbitrarily long
wavelengths it is necessary that 
\be
\lambda \geqslant 0
\ee
Otherwise, these wave packets will get $Im\, \omega\neq 0$.
\par
The strategy is now to study the radial equation of the set (\ref{eqa})
and check whether there is an aceptable set of solutions with $\lambda \geqslant 0$.
\par
%%%%%%%%%%%%%%%%%%%%%%%%%%%%%%%%%%%%%%%%%%%%%%%%%%%%%%%%%%%%%%%%%%%%%%%%%%%%%%%%%%%%%%%
\subsubsection{Asymptotic behavior}
%%%%%%%%%%%%%%%%%%%%%%%%%%%%%%%%%%%%%%%%%%%%%%%%%%%%%%%%%%%%%%%%%%%%%%%%%%%
Our noncritical solution is
generically characterized by
\be
a'= c_1 a^{1+\frac{\sqrt{d}}{2}}+ c_2 a^{1-\frac{\sqrt{d}}{2}}
\ee
and $c_1 c_2 <0$. 
Specially interesting is the behavior of the preceding equation (\ref{eqa})
in the neighborhood of the singularity ($a\rightarrow 0$) or at $r\rightarrow
\infty $ ($a\rightarrow 0$).
\par
When $r\rightarrow \infty$ the behaviour is universal, (for any $D<26$, i.e.
$c_1c_2<0$), the scale factor $a$ remains bounded, 
and equivalent to the 
behaviour when $x\sim 0$ (a regular singular point) of:
\be
T''- T'= \frac{C}{x^2}T
\ee
where $C\equiv \frac{m^2-\frac{\lambda}{\bar{a}}}{- c_2 c_1\, d} $. The behavior of the solution depends critically as to whether $1-d-\lambda > 0$, in which
case there are two real powers $x^{\a_{\pm}}$. When the equality is 
saturated, that is, $d=1-\lambda$  there is a singular 
solution, $ x^{1/2}log x $ (which however vaniches in the limit 
$x\rightarrow 0$). When $1-d-\lambda<0$ there is an oscillatory
behaviour with decreasing amplitude $\sim x^{1/2}$.
\par
In the oposite situation, that $c_1c_2>0$, $a\rightarrow \infty$, and the 
behaviour of the wave equation is then equivalent to the behaviour when
$x\sim 0$ of:
\be
\frac{d^2  T}{dx^2}+\frac{1}{x}\frac{d T}{d x}=\frac{m^2}{c_1^2}x^{-2+\sqrt{d}}T
\ee

\par
When $a\rightarrow 0$ i.e., a neighborhood of the singularity (and this 
is now independent of the signs of the constants $c_1c_2$) the solutions of equation (\ref{eqa}) behave as:

\begin{itemize}
\item $d>1$. The behaviour of the solutions is, for $\lambda \neq 0$:
\bea
T_1&&\sim 1 - \frac{\lambda}{c_2^2 (\sqrt(d)-1)^2}a^{\sqrt{d}-1}\nonumber\\
T_2&&\sim \log{     a^  {    \frac {\sqrt{d}-1}{2}    }       }
\eea
And for $\lambda =0$:
\bea
T_1&&\sim 1 + \frac{m^2}{c_2^2 d}a^{\sqrt{d}}\nonumber\\
T_2&&\sim \log{a^{\frac{\sqrt{d}}{2}}}
\eea 
\item $d=1$. In this case the solutions are replaced by:\\
When $\lambda \neq 0$:
\be
T_{\pm} \sim  a^{\pm \sqrt{-\lambda /c_ 2^2}}\nonumber\\
\ee2
And for $\lambda=0$:
\bea
T_1&&\sim 1- \frac{m^2}{c_2^2} a\nonumber\\
T_2&&\sim \log{a}
\eea
\end{itemize}
In summary for $D>2$ all solutions in the region $r \rightarrow \infty$
tend to zero oscillating. At the origin conservation of energy and
normalizability force us to rule out the logaritmic behavior but they allow us to
have a tachyon tending to a non vanishing constant. Next we will consider
the exact solution in the special case of $\lambda = 0$.
%%%%%%%%%%%%%%%%%%%%%%%%%%%%%%%%%%%%%%%%%%%%%%%%%%%%%%%%%%%%%%%%%%%%%%%%%%%%%
\subsubsection{Exact Solution}
%%%%%%%%%%%%%%%%%%%%%%%%%%%%%%%%%%%%%%%%%%%%%%%%%%%%%%%%%%%%%%%%%%%%%%%
Denoting $q= \frac{\sqrt{d}}{2}$ and changing variables
\bea\label{var}
a&&=e^{z}\nonumber\\
 x&&= th(qz)
\eea
equation (\ref{eqa}) for the tachyon (with $\lambda=0$) becomes
\be
(1-x^{2})T'' -2xT' + \frac{6}{(25-d)x^{2}} T =0
\ee
for $T' = \frac{dT}{dx}$.
This is the Riemann equation \cite{abramowitz} with three singular points at $x=1,-1,0$. The general
solution is
\be
T = P\left( \begin{array}{cccc}
-1 & 1 &  0 & \\
0 & 0 & \gamma & x \\
0 & 0 & 1-\gamma & 
\end{array} \right) =\left( \frac{x}{x-1} \right) ^\gamma  
 P\left( \begin{array}{cccc}
0 & \infty &  1 & \\
0 & \gamma & 0 & \frac{1+x}{1-x} \\
0 & \gamma & 1-2 \gamma & 
\end{array} \right)
\ee
with 
\be
\gamma= \frac{1+\sqrt{\frac{1-d}{25-d}}}  {2}
\ee
We have two linearly independent solutions to be denoted $T_{1}$ and $T_{2}$. Around the singular points they behave
like:
\be
T^{(1)}_{x=-1}=c_1 \left( \frac{x}{x-1} \right) ^{\gamma} F[\gamma,\gamma,1; \frac{1+x}{1-x}] \sim cte
\ee
\be
T^{(1)}_{x=0}=c_1 \left( \frac{x}{x-1} \right) ^{\gamma} F[\gamma,\gamma,2 \gamma; \frac{-2x}{1-x}] \sim x^{\gamma} 
\ee
and 
\be
T^{(2)}_{x=-1} = c_2 T^{(1)}_{x=-1} \log \left( \frac{1+x}{1-x} \right) +  c_2S(x) \sim log(1+x)\\
\ee
\be
T^{(2)}_{x=0}=c_2 \left( \frac{x}{x-1} \right) ^{\gamma} \left( \frac{-2x}{1-x} \right) ^{1-2 \gamma} F[1-\gamma,1-\gamma,-2 \gamma; \frac{-2x}{1-x}] \sim x^{1- \gamma}
\ee
\\
Where $S(x)$ is given by:
\be
S(x) = \sum_{n}^{\infty}\left( \frac{(\gamma)_n ^2 }{(n!)^2} \left( \frac{1+x}{1-x} \right) ^n \left( 2\psi(\gamma +n) -2\psi(\gamma) -2\psi(n+1) + 2 \psi(1) \right) \right)
\ee
(where $\psi$ is the digamma function). Notice that $S(x) \rightarrow 0$ when $x\rightarrow -1$.

The only one compatible with energy conservation at the origin is 
$T_{1}$ that precisely looks like a kink interpolating between $<T> = 0$ and $<T> = cte$
(see figure 2). This concludes our discussion of the exact tachyon solution.

%%%%%%%%%%%%%%%%%%%%%%%%%%%%%%%%%%%%%%%%%%%%%%%%%%%%%%%%%%%%%%%%%%%%%%%%%%%%
\section{Discussion and Final Comments}
%%%%%%%%%%%%%%%%%%%%%%%%%%%%%%%%%%%%%%%%%%%%%%%%%%%%%%%%%%%%%%%%%%%%%%%%%%%%

Starting with Sen's conjecture \cite{sen} on open tachyon condensation, several candidates have
recently appeared for the open tachyon potential (cf.\cite{kutasov}, \cite{kutasov1}), and even
for the closed tachyon potential (cf. \cite{tseytlin}).

\begin{figure}[h]
\begin{center}
\leavevmode
\epsfxsize=4cm
\epsffile{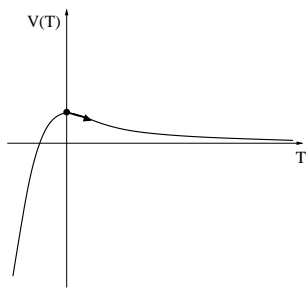}
\caption{\it Open string Tachyon potential}
\label{fig1}
\end{center}
\end{figure}         

 To be specific, in the open string case tachyon
condensation can be understood on the basis of a tachyon potential
of type $e^{-T}(1+T)$ (see Fig 3) 
derived from Witten's \cite{witten} \cite{gerasimov} 
background independent open string field theory . In the case of
the closed string tachyon the situation is still unclear, however as we have already said, there
is a recent suggestion in \cite{tseytlin} ,based on $\sigma$ - model analysis,
of a closed string tachyon potential of type $-T^{2}e^{-T}$ (see Fig 4).

\begin{figure}[h]
\begin{center}
\leavevmode
\epsfxsize=4cm
\epsffile{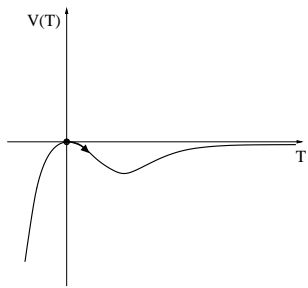}
\caption{\it Closed string tachyon potential $-T^{2}e^{-T}$. The arrow indicates the direction of condensation}        
\label{fig1}
\end{center}
\end{figure}

It is of course tempting to interpret the tachyon solution $T_{1}$ ,
we have just derived for
the new family of backgrounds  presented in this paper, as 
describing precisely the tachyon condensation for this particular closed string
tachyon potential. A question that remains open is the dynamical meaning
of the singularity, probably related to the particular way in which
this background solution encodes the condensation of degrees of freedom.
Incidentally, were this solution  to be interpreted as a confining background,
the static potential between heavy sources as derived from the lowest order Wilson loop
computation yields a linear behavior \cite{manjarin}.

%%%%%%%%%%%%%%%%%%%%%%%%%%%%%%%%%%%%%%%%%%%%%%%%%%%%%%%%%%%%%%%%%%%%
\section*{Acknowledgments}
We are indebted to Pedro Resco for much help with the computations.
Useful correspondence with Arkady Tseytlin is gratefully acknowledged.
This work ~~has been partially supported by the
European Union TMR program FMRX-CT96-0012 {\sl Integrability,
  Non-perturbative Effects, and Symmetry in Quantum Field Theory} and
by the Spanish grant AEN96-1655.  The work of E.A.~has also been
supported by the European Union TMR program ERBFMRX-CT96-0090 {\sl 
Beyond the Standard model} 
 and  the Spanish grant  AEN96-1664. L.H. is supported by the spanish 
predoctoral grant AP99 43367460.

%%%%%%%%%%%%%%%%%%%%%%%%%%%%%%%%%%%%%%%%%%%%%%%%%%%%%%%%%%%%%%%%%%%%%%

%%%%%%%%%%%%%%%%%%%%%%%%%%%%%%%%%%%%%%%%%%%%%%%%%%%%%%%%%%%%%%%%%%%%%%


\begin{thebibliography}{99}



\bibitem{abramowitz}M. Abramowitz and I. Stegun,
{\em Handbook of Mathematical Functions}, (Dover University Press).


\bibitem{ag}E. Alvarez and C. G\'omez,
{\em The Confining String from the Soft Dilaton theorem}
Nucl.\ Phys.\  {\bf B566} (2000) 363
{\tt hep-th/9907158}

\bibitem{manjarin} E. Alvarez and  J.J. Manjar\'{\i}n,
{\em Static Gauge Potential from Non-critical Strings}, to appear.
\bibitem{bf} P. Breitenlohner and D. Freedman,
{\em Stability in Gauge Extended Supergravity},
Ann. Phys. 144 (1982),249.


\bibitem{englert}
A.~Casher, F.~Englert, H.~Nicolai and A.~Taormina,
{\em Consistent Superstrings As Solutions Of The D = 26 
Bosonic String Theory},
Phys.\ Lett.\  {\bf B162} (1985) 121.
%%CITATION = PHLTA,B162,121;%%

\bibitem{gerasimov} Anton A. Gerasimov, Samson L. Shatashvili,
{\em On Exact Tachyon Potential in Open String Field Theory},
 JHEP 0010 (2000) 034, {\tt hep-th/0009103}.\\
{\em Stringy Higgs Mechanism and the Fate of Open Strings},
{\tt hep-th/0011009}. 
                                     

\bibitem{kutasov1}
J.~A.~Harvey, D.~Kutasov and E.~J.~Martinec,
{\em On the relevance of tachyons},
{\tt hep-th/0003101}.


\bibitem{kutasov}
D.~Kutasov, M.~Marino and G.~Moore,
{\em Some exact results on tachyon condensation in string field theory},
{\tt hep-th/0009148}.




\bibitem{myers} R. Myers, 
  {\em New Dimensions for old strings},
 Phys.Lett.B199:371,1987 


\bibitem{polchinski}
J. Polchinski,
                  {\em String Theory},
                   Cambridge University Press.
%\bibitem{bender}C.M.Bender and S.A.Orszag,
%{\em Advanced Mathematical Methods for Scientists and Engineers},
% Mc Graw Hill.


\bibitem{polyakov}
A.~M.~Polyakov,
{\em The wall of the cave},
Int.\ J.\ Mod.\ Phys.\  {\bf A14} (1999) 645
{\tt hep-th/9809057}.

\bibitem{sen}
A.~Sen,
{\em Universality of the tachyon potential},
JHEP {\bf 9912} (1999) 027
{\tt hep-th/9911116}.






\bibitem{tseytlin}
A.A.Tseytlin, 
{\em Sigma Model Approach to string theory effective actions with tachyons},
{\tt hep-th 0011033}



\bibitem{witten}
E.~Witten,
{\em On background independent open string field theory},
Phys.\ Rev.\  {\bf D46} (1992) 5467
{\tt hep-th/9208027}.\\
{\em Some computations in background independent off-shell string theory},
Phys.\ Rev.\  {\bf D47} (1993) 3405
{\tt hep-th/9210065}.





\end{thebibliography}
\end{document}